\documentclass[aps,prd,twocolumn,nofootinbib,floatfix,superscriptaddress]{revtex4-1}
\usepackage{setspace}
\usepackage{graphicx}
\usepackage{dcolumn}
\usepackage{multirow}
\usepackage{subfigure}
\usepackage{times,mathptm}
\usepackage{float}
\usepackage{color}
\usepackage{amsmath,amsfonts}
\usepackage{mathptmx}
\usepackage{mathrsfs}
\usepackage{bbm}
\usepackage{bm}
\usepackage{xfrac}

\newcommand{\beq}{\begin{equation}}
\newcommand{\eeq}{\end{equation}} 
\newcommand{\bea}{\begin{eqnarray}}
\newcommand{\eea}{\end{eqnarray}}

\newcommand{\J}{{\cal J}}

\renewcommand{\d}{\delta}

\newcommand{\CD}{{\cal D}}
\newcommand{\tA}{\widetilde{A}}

\renewcommand{\b}{\beta}
\renewcommand{\a}{\alpha}

\newcommand{\tr}{\text{Tr}}

\newcommand{\tp}{\widetilde{\phi}}

\newcommand{\tn}{\tilde{n}}
\newcommand{\tm}{\tilde{m}}
\renewcommand{\o}{\omega}

\newcommand{\vx}{{\vec{x}}}
\newcommand{\vy}{{\vec{y}}}

\newcommand{\vk}{{\vec{k}}}

\newcommand{\q}{\overline{q}}
\newcommand{\g}{\gamma}
\renewcommand{\r}{\rho}

\newcommand{\s}{\sigma}

\newcommand{\D}{\Delta}
\newcommand{\tD}{\widetilde{D}}

\newcommand{\M}{{\cal M}}
\renewcommand{\th}{\theta}

\newcommand{\oh}{\frac{1}{2}}

\newcommand{\dg}{\dagger}
\newcommand{\non}{\nonumber}

\newcommand{\rf}[1]{(\ref{#1})}
\newcommand{\ra}{\rightarrow}

\renewcommand{\vec}[1]{\bm #1}

\usepackage{ulem}

\begin{document}

\title{Coulomb string tension, asymptotic string tension, and the gluon chain} 

\bigskip
\bigskip

\author{Jeff Greensite}
\affiliation{Physics and Astronomy Department, San Francisco State
University,   \\ San Francisco, CA~94132, USA}
\bigskip
\author{Adam P. Szczepaniak}
\affiliation{Department of Physics, Indiana University, Bloomington, IN 47405, USA}
\affiliation{Center for Exploration of Energy and Matter, Indiana University, Bloomington, IN 47403, USA}
\affiliation{Theory Center, Thomas Jefferson National Accelerator Facility, 12000 Jefferson Avenue, Newport News, VA 23606, USA}
\date{\today}
\vspace{60pt}
\begin{abstract}

\singlespacing
 
We compute, via numerical simulations, the non-perturbative Coulomb potential of pure SU(3) gauge theory in Coulomb gauge.  We find that that the Coulomb potential scales nicely in accordance with asymptotic freedom, that the Coulomb potential is linear in the infrared, and that the Coulomb string tension is about four times larger than the asymptotic string tension.  We explain how it is possible that the asymptotic string tension can be lower than the Coulomb string tension by a factor of four.

\end{abstract}

\maketitle

\singlespacing
\section{\label{intro}Introduction}

   In this article we will report on  a  lattice Monte Carlo computation of the long-range instantaneous Coulomb potential between static color sources in SU(3) pure gauge theory.  Note that by ``Coulomb potential'' we are not referring to the perturbative $1/r$ expression, but rather to the expectation value of the full non-local term in the Coulomb-gauge Hamiltonian associated with Coulomb energy.  We will find that this potential is linear at large separations, that the potential scales as expected with lattice coupling, and that the Coulomb string tension $\s_c$ is about four times larger than the accepted value $\s=$ (440 MeV)${}^2$ for the asymptotic string tension.  Since gluons cannot possibly screen a color source in the fundamental representation, the obvious question is: what mechanism can reduce the Coulomb string tension by a factor of four, while retaining the linearity of the potential?  We will try to answer this question in the context of a model in which the QCD flux tube is pictured as a superposition of states containing different numbers of constituent gluons, held together by Coulombic interactions, and arranged roughly in a chain between the static sources.
   
    Let us first be a little more explicit about what is meant by the term ``Coulomb potential.''  It is really the interaction energy of a particular physical state, which is simply expressed in Coulomb gauge as a pair of static quark-antiquark operators, separated by a spatial distance $R$, operating on the (non-perturbative) vacuum state 
\beq
           |0\rangle_{\q q} = \q^{\dag}_i(0) q^{\dag}_i (R) |0\rangle_{\text{vac}} \ ,
\label{0g}
\eeq 
where 
\beq
          \Psi_0[A] = \langle A |0\rangle_{\text{vac}}
\eeq
is the true vacuum wavefunctional.\footnote{Ideas about the form of this wavefunctional go back a long way, cf. 
\cite{Greensite:2011pj} and references therein.  Those ideas will not be needed, however, in the present investigation.}  The energy expectation value of such a state is given by the logarithmic time derivative
\beq
          V(R) = - \lim_{t\ra 0} {d \over dt} \log \big\{ {}_{\q q}\langle 0| e^{-Ht}  |0\rangle_{\q q} \big\} \ ,
\eeq
where $H=H_{glue}+H_{coul}$ is the Coulomb gauge Hamiltonian for a pair of static quark-antiquark  sources,  
\begin{eqnarray}
        H_{glue} = \oh \int d^3x\;( \J^{-\oh}\vec{E}^{{\rm tr},a} {\cal J}
        \cdot \vec{E}^{{\rm tr},a} \J^{-\oh} + \vec{B}^a \cdot \vec{B}^a),
\non \\
        H_{coul} = \oh \int d^3x d^3y\;\J^{-\oh}\r^a(x) \J
                K^{ab}(x,y;A) \r^b(y) \J^{-\oh},
\non \\
\end{eqnarray}
with 
\begin{eqnarray}
        K^{ab}(x,y;A) &=& \left[ \M^{-1}
        (-\nabla^2) \M^{-1} \right]^{ab}_{xy},
\non \\
        \r^a &=& \r_q^a  + \r^a_{\bar q}   + \rho^a_g  ,
\non \\
         \M &=& -\nabla \cdot \CD(A) ~~,~~ \J = \det[\M].
\end{eqnarray}
Here  $\r^a_q(x) = g  q^\dag_i(x) T^a_{ij} q_j(x)$, $\r^a_{\bar q}(x) = g  \bar q_i(x) T^a_{ij} \bar q^{\dag}_j(x)$ and  $\rho^a_g(x) = - g f^{abc} A^b_k(x) E^c_k(x)$
 are the  charge density of quarks, antiquarks and gluons, respectively, and $\CD_k(A)$ is the covariant derivative.  Since we are taking the $t\ra 0$ limit,  the contribution from connected diagrams to the energy expectation value 
comes from the non-local Coulomb term proportional to $K(\vx-\vy;A)$, which contributes to both the quark self-energies
and to an $R$-dependent Coulomb interaction.   As Dirac indices and quark kinetic energies are not relevant to our study, it is sufficient to compute, in a Euclidean action formulation, the logarithmic time derivative of a correlator of short timelike Wilson lines
\beq
           V(R) = - \lim_{t\ra 0} {d \over dt} \log  \big\langle \tr[L_t({\bf 0}) L_t^\dg({\bf R})] \big\rangle \ ,
\eeq
where
\beq
              L_t(\vx) \equiv T\exp\left[ig\int_0^t dt A_0(\vx,t) \right] \ .
\eeq
Again it should be stressed that $V(R)$ contains both an $R$-dependent interaction, and an $R$-independent Coulomb
self energy.  On the lattice, for SU(3) gauge theory, this becomes a correlator of timelike link operators on timeslice $t=0$:
\bea
          V(R) &=&  \lim_{\b \ra \infty} \left({V_L(R_L,\b) \over a(\b)} \right) \ ,
\non \\
           V_L(R,\b)  &=& - \log \big\langle {1\over 3} \tr U_0({\bf 0},0) U^\dg_0({\bf R},0) \big\rangle \ ,
\eea
where $R_L$ is the quark-antiquark separation in lattice units, $R=R_L a(\b)$, and $a(\b)$ is the lattice spacing (same in all directions) at Wilson lattice coupling $\b$.  On a periodic lattice one can average over different timeslices.  

    This method for computing the instantaneous Coulomb potential was first suggested in ref.\ \cite{Greensite:2003xf,*Greensite:2004ke}, and the calculation was carried out for the SU(2) gauge group.  There is another possible 
approach, adopted in ref.\ \cite{Burgio:2012bk} for SU(2) and in ref.\ \cite{Voigt:2008rr} for SU(3) gauge groups, which is to directly compute the expectation value of the operator $K(\vx-\vy,A)$, Fourier transformed to momentum space.  This involves inverting the Faddeev-Popov operator $\M=-\nabla \cdot \CD(A)$, and looking for a plateau 
in $k^4 V(k)$.\footnote{These authors find a Coulomb string tension which is 2.2 \cite{Burgio:2012bk} or 1.6 
\cite{Voigt:2008rr} times the asymptotic string tension.  Our result, reported in the next section, is substantially
higher than those values.} We prefer to use the original approach of \cite{Greensite:2003xf,*Greensite:2004ke} which, we believe, provides better evidence of the linearity of the Coulomb potential.   

\section{The instantaneous Coulomb potential} 

\begin{figure}[t!]
\centerline{\scalebox{0.6}{\includegraphics{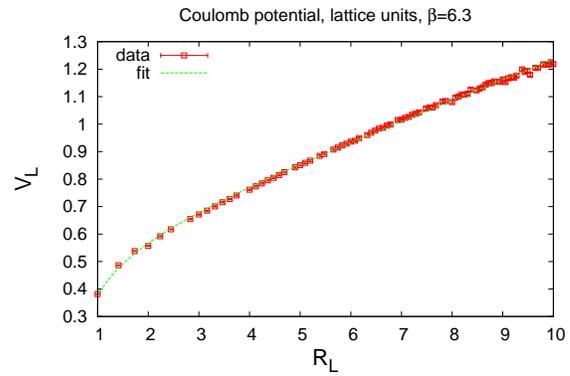}}}
\caption{The instantaneous Coulomb potential in lattice units at $\b=6.3$.  This data includes a Coulomb self-energy term
for the static sources.  The solid line is a fit to eq.\ \rf{VL}.  Error bars are comparable to but smaller than the symbol size.}
\label{pot63}
\end{figure}

      We have calculated the instantaneous Coulomb potential by the method just described on a $24^4$ hypercubic lattice
in SU(3) pure gauge theory with a standard Wilson action and lattice coupling $\b$ in the range $\b \in [5.9,6.4]$.  The method of Fourier acceleration is used for Coulomb gauge fixing \cite{Giusti:2001xf,*Davies:1987vs}.    An example of the data for $V_L(R,\b)$, at $\b=6.3$, is shown in Fig.\ \ref{pot63},
together with a best fit to the functional form
\beq
           V_L(R_L,\b) = \s_L(\b) R_L - {\g(\b) \over R_L} + c(\b) \ .
\label{VL}
\eeq
Note that the data includes off-axis separations.  Only the data point at $R_L=0$ is excluded in fitting the data.  
The constant $c$ is the self-energy and $\s_L$ is the Coulomb string tension, both in lattice units.  The interaction energy is obtained by subtracting the self energy $c(\b)$ from the data, i.e.\ $V_L^{int}(R_L,\b) = V_L(R_L,\b)-c(\b)$.  To convert everything to physical units we divide both sides by the lattice spacing $a(\b)$ and multiply by a conversion factor
($0.197$ Gev-fm = 1) taking inverse fm to GeV,
\beq
          V^{int}(R,\b) = \s_{c}(\b) R - (0.197~\text{GeV-fm}) {\g(\b) \over R} \ ,
\eeq
where $V^{int}$ is in GeV, $R=R_L a(\b)$ is in fm, and $\s_c$ is the Coulomb string tension in units of Gev/fm.
As $\b \ra \infty$, the interaction energy in physical units  (and consequently $\s_c$ and $\g$) should tend to a finite
limit.  For $a(\b)$ we use the Necco-Sommer formula \cite{Necco:2001xg}  
\bea
a &=& r_0 \exp\Bigl(-1.6804-1.7331 (\b-6) 
\non \\
& & \qquad + 0.7849(\b-6)^2  - 0.4428(\b-6)^3 \Bigr) \ ,
\label{scaling}
\eea
with $r_0=0.5$ fm, for every lattice spacing in the range \newline
$\b \in [5.9,6.4]$.   

\begin{figure}[t]  
 \includegraphics[scale=0.6]{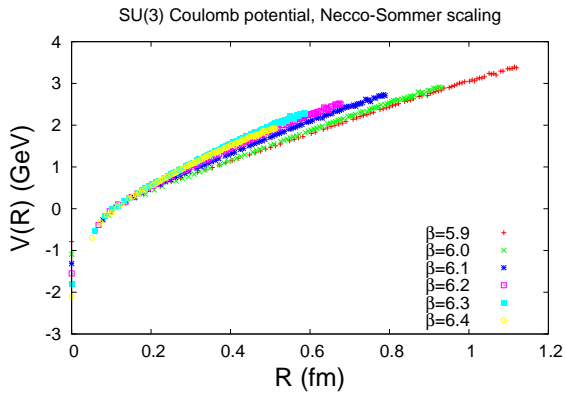}
\caption{The instantaneous Coulomb potential in physical units, for a range of lattice couplings $\b$, with self-energies subtracted as explained in the text.}
\label{potential}
\end{figure}

   The result for the Coulomb potential in physical units is shown in Fig.\  \ref{potential}.  With the self-energy term $c(\b)/a(\b)$ removed, the data for $V^{int}(R,\b)$ seems to converge nicely to a limiting curve as $\b$ increases. 

\begin{figure}[t]
 \includegraphics[scale=0.6]{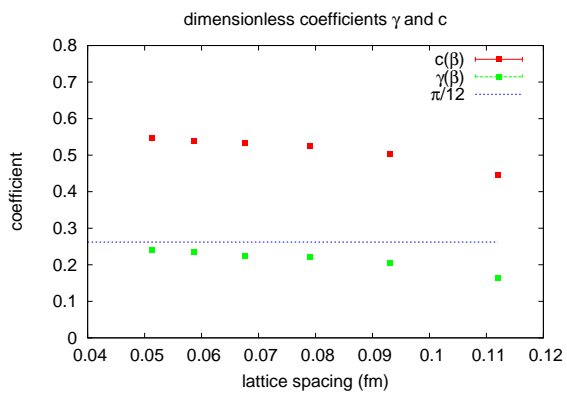}
\caption{The self-energy term $c(\b)$, and the coefficient $\g(\b)$ of the $1/R_L$ term in the instantaneous
Coulomb potential, derived from a fit to \rf{VL}, vs.\  lattice spacing $a(\b)$. The flat line at $\pi/12$ indicates the value of the coefficient of the L\"uscher term.  Error bars are smaller than the symbol size.}
\label{gc}
\end{figure}

    In Fig.\ \ref{gc} we show our data for the dimensionless parameters $\g(\b)$ and $c(\b)$, plotted vs.\  lattice spacing $a(\b)$.  Both of them appear to be
converging to a finite limit as $\b \ra \infty, a \ra 0$.  What is curious, however, is that the limit for $\g(\b)$ might very well be consistent with the coefficient of the L\"uscher term, i.e.
\beq
           \g = {\pi \over12}=0.262 \ .
\eeq
It is hard to know whether or not this is a coincidence.  The Coulombic field of a $q\q$ pair, while confining, is nonetheless
extended.  There is no particular reason to believe that it is collimated into a flux tube, or has string-like properties.  At present we cannot explain why $\g$ would have this particular limit.

\begin{figure}[t]
 \includegraphics[scale=0.6]{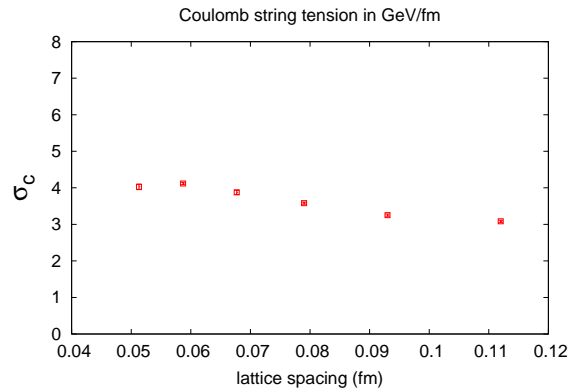}
\caption{The Coulomb string tension $\s_c$ in physical units (Gev/fm) vs.\  lattice spacing $a(\b)$.  Error bars are smaller than the symbol size.}
\label{tension}
\end{figure} 

  The Coulomb string tension determined at each $\b$, vs.\ lattice spacing $a(\b)$, is shown in Fig.\ \ref{tension}.  The value of the string tension at the smallest lattice spacing is almost within errorbars of the next two data points, which suggests that the data may have converged to the asymptotic value.  From the last data point at $\b=6.4,~a=0.051$ fm, we estimate the Coulomb string tension to be $\s_c \approx 4.03(8) ~ \mbox{GeV/fm}$, or in other units $\s_c = (891 \pm 9 ~ \mbox{MeV})^2$, to be compared to the accepted value of $\s = (440 ~ \mbox{MeV})^2$ for the asymptotic string tension.
These values differ by more than a factor of four, and a discrepancy of this size cries out for an explanation.  How can the
asymptotic string tension be so much smaller than the ``natural'' value obtained from the instantaneous Coulomb potential?

\section{\label{chains}Gluon chains}

   The starting point is that since the instantaneous Coulomb potential is the interaction potential of a certain physical
state, namely \rf{0g}, and the energy of that state ($\approx \s_c R$) is far larger than $\s R$ for large $\q q$ separations, 
it must be that \rf{0g} is not the minimal energy $\q q$ state.  So, what {\it is} the minimal energy state, and how can the 
string tension in that state be so much lower than $\s_c$? 

   The original idea of the {\it gluon chain model} \cite{Greensite:2001nx} was as follows:  Suppose that as a quark antiquark pair separate, the interaction energy eventually starts to rise at a rate faster than linear in the separation.  At some point, call it 
$R=R_c$, it becomes energetically favorable to insert a gluon between the quark antiquark pair to reduce the
separation of color charges to roughly $R_c/2$.  As the quark and antiquark continue to separate, eventually it becomes
favorable to insert a second gluon, and so on, so that no matter what the separation of the quark and antiquark, the
average separation of color charges is no more than $R_c$.  Let us suppose that for very large quark separation $R$,
the average distance between gluons is $R_{av}$, so there will be approximately $N=R/R_{av}$ gluons ordered in a chain
between the quark and antiquark. Let $E(R_{av})$ denote the kinetic energy plus the share of Coulomb interaction energy
carried by each gluon.  Then the total energy of the chain is $V(R) = NE(R_{av}) = \s R$, where $\s=E(R_{av})/R_{av}$.

  Our numerical investigations, and prior studies \cite{Greensite:2003xf,*Greensite:2004ke}, \cite{Burgio:2012bk}, \cite{Voigt:2008rr},  show that this simple picture is untenable, because the increase in Coulombic interaction energy with separation is asymptotically linear.  Inserting more gluons between the quarks not only increases the energy of the state by the kinetic energy of each gluon, but also increases the Coulombic energy.  If the gluons were arranged exactly along a line between the quarks, and the interaction energy between neighboring gluons is $\s_c$ times gluon separation, then the total Coulomb
interaction energy of the chain is $\s_c R$, no matter how many gluons are in the chain.  The inevitable fluctuations
in gluon position in directions transverse to the line defined by the $\q q$ pair will only increase this interaction energy.  It would then appear that the lowest energy state is the zero gluon state $|0\rangle_{\q q}$, and we have already seen that the string
tension of this state is four times larger than the asymptotic string tension.

   However, this conclusion ignores the fact that a state with $n$-constituent gluons is not an eigenstate of the Coulomb gauge Hamiltonian.  There will obviously be matrix elements of the Hamiltonian connecting states with different numbers of constituent gluons, and it is interesting to consider, even at a very crude and qualitative level, what the effect of those off-diagonal elements might be.\footnote{For some earlier discussions of how constituent gluons might lower the Coulomb string tension, see \cite{Ostrander:2012kz,*Szczepaniak:2005xi} and \cite{Greensite:2009mi}.}

    Let us define the operator
\beq
          \tA_i(\vx) = \int {d^3k \over (2\pi)^3} ~  \sqrt{2\o(k)} A_{i}(k) e^{i\vk \cdot \vx} \ ,
\eeq
where $A_i(k)$ is the Fourier transform of the position-space operator $A_i(x) = A^a_i(x) T^a$, and $\o(k)$ depends on the
transverse gluon propagator.  For a free massless field, $\o(k)=|\vk|$.
Then we define an $n$ constituent-gluon state to be a state of the form
\beq
       |n\rangle_{\q q} =  \q^{\dag}_i ({\bf 0}) \psi_n^{ij}[A;f] q_j^{\dag}({\bf R}) |0\rangle_{vac} \ ,
\label{state1}
\eeq
where       
\bea
 \psi_n^{ij}[A;f] &=&        
       \int d^3x_1 d^3x_2 ... d^3x_n ~ f^{(n)}_{k_1 k_2 ... k_n}(\vx_1,\vx_2,...,\vx_n)
\non \\
  & &      \times [\tA_{k_1}(\vx_1)
       \tA_{k_2}(\vx_2)... \tA_{k_n}(\vx_n)]_{ij}\ .
\label{state2}
\eea
Color matrix indices are contracted such that the $n$-gluon state is invariant with respect to global color rotations, which are consistent with the Coulomb gauge condition. We can suppose that the quark and the antiquark lie a distance $R$ apart along the $z$-axis.  If the function $f$ is such that it is large when the ordering of gluon fields along the $z$ axis corresponds to
their color ordering; i.e.\ when $0 < z_1 < z_2 < ... < R$, and is strongly suppressed when this ordering is violated, then we will refer to $ |n\rangle_{\q q}$ as a  ``gluon chain'' state.  Moreover, for the reason mentioned above, in order to bound the Coulomb energy the fluctuations in gluon position transverse to the $z$-axis should not be too large, so that the $n$ gluon operators are contained in a roughly cylindrical region of some kind.  A simple example of a function with these properties
is
\bea
\lefteqn{f^{(n)}_{k_1 k_2 ... k_n}(\vx_1,\vx_2,...,\vx_n)}
\non \\
& & \qquad = \d_{k_1 3} \d_{k_2 3}...\d_{k_n 3}   
\non \\
& & \qquad \qquad \times \th(z_1) \th(z_2-z_1) \th(z_3-z_2) \cdot \cdot \cdot \th(R-z_n)
\non \\
& & \qquad \qquad  \times \exp\left[-{1\over 8} a^2 \sum_{i=1}^n (x_i^2 + y_i^2)\right] \ .
\label{f}
\eea
The constant $a$ can be regarded as a variational parameter.  This is not necessarily the optimal choice for $f^{(n)}$, and of course one can consider other more complicated functions containing many parameters.  But it will serve to illustrate what we
have in mind.

    Having settled on some choice for the $f^{(n)}$, we can in principle orthogonalize and normalize a finite set of $N$ states 
$\{|n\rangle_{\q q}, ~n=0,1...,N\}$ by, e.g., the Gram-Schmidt procedure.  Let us denote the resulting set of states 
$\{|\tn\rangle,~n=0,1...,N\}$, with Hamiltonian matrix elements
\beq
           H_{nm} = \langle \tn |H|\tm \rangle \ .
\label{Hnm}
\eeq
The prescription is then to diagonalize this finite matrix.  The lowest eigenvalue provides us with an estimate of the
energy of the $\q q$ state. The Hamiltonian matrix elements can be determined from the finite-time amplitude
\beq
          T_{nm}(t) =    {}_{\q q}\langle n| e^{-Ht} | |m\rangle_{\q q} \ ,
\eeq
or, stripping away irrelevant Dirac indices, the correlator
\beq
          T_{nm}(t) =   \big\langle \tr[L_t({\bf 0}) \psi_n[A(\vx,0);f]  L_t^\dg({\bf R}) \psi_m^{\dg}[A(\vx,t);f] ] \big\rangle \ .
\eeq
$T_{nm}(0)$ gives us the information required to construct a set of normalized, orthogonal states
\beq
         |\tn\rangle = \sum_{m=0}^N C_{nm} |m \rangle_{\q q} \ ,
\eeq
while the time derivative
\beq
      -   \lim_{t\ra 0} {d \over dt} T_{nm}(t) =  {}_{\q q}\langle n| H |m\rangle_{\q q}
\eeq
contains the rest of the information required to construct $H_{nm}$, i.e.
\bea
         H_{nm} = \sum_j \sum_k C^*_{nj} \left\{- \lim_{t\ra 0} {d \over dt} T_{jk}(t) \right\} C_{mk} \ .
\eea

\begin{figure*}[t]
\subfigure[]  
{   
 \label{kinetic}
 \includegraphics[scale=0.32]{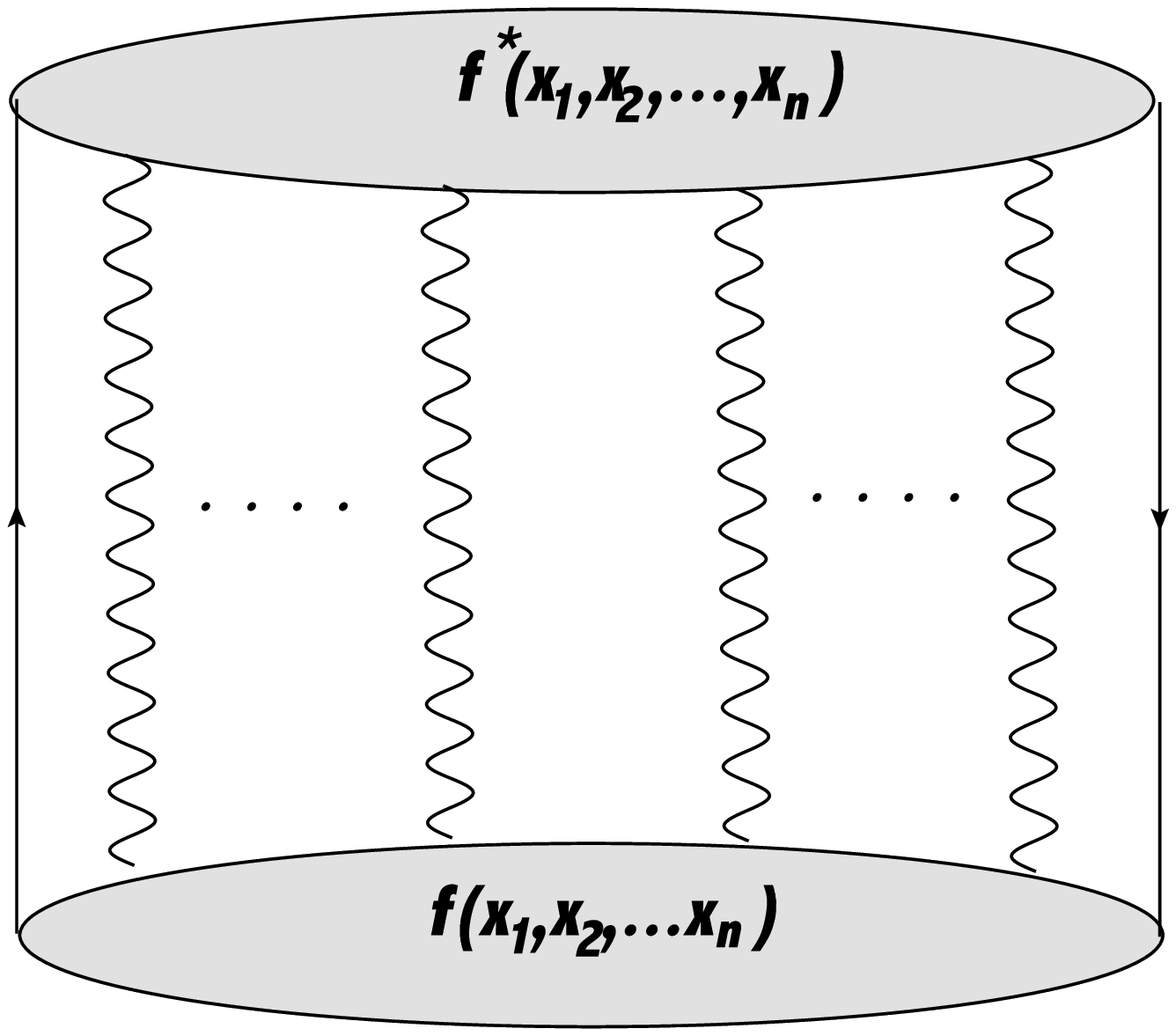}
}
\subfigure[]  
{   
 \label{coulomb}
 \includegraphics[scale=0.32]{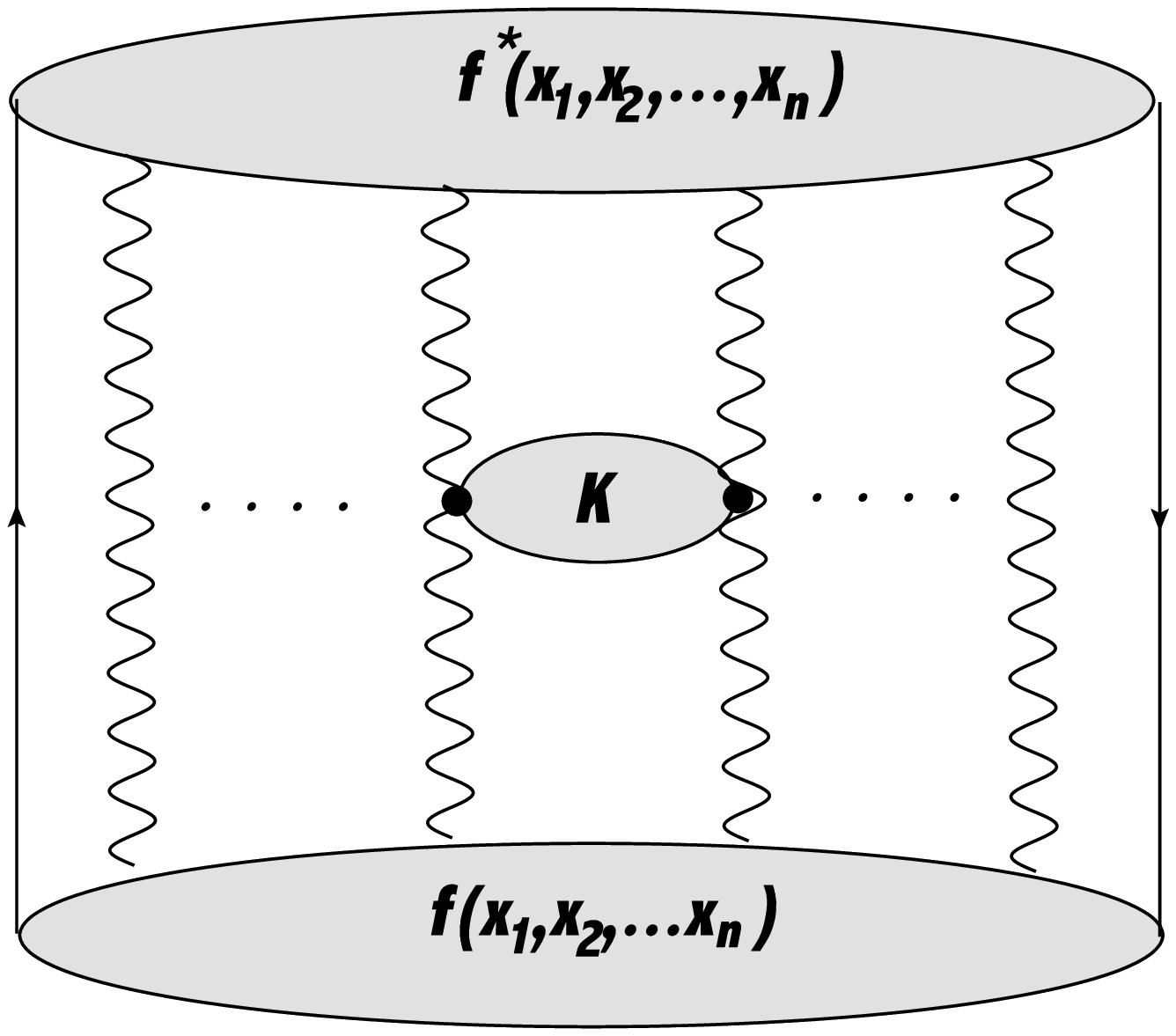}
}
\subfigure[]  
{   
 \label{offdiagonal}
 \includegraphics[scale=0.32]{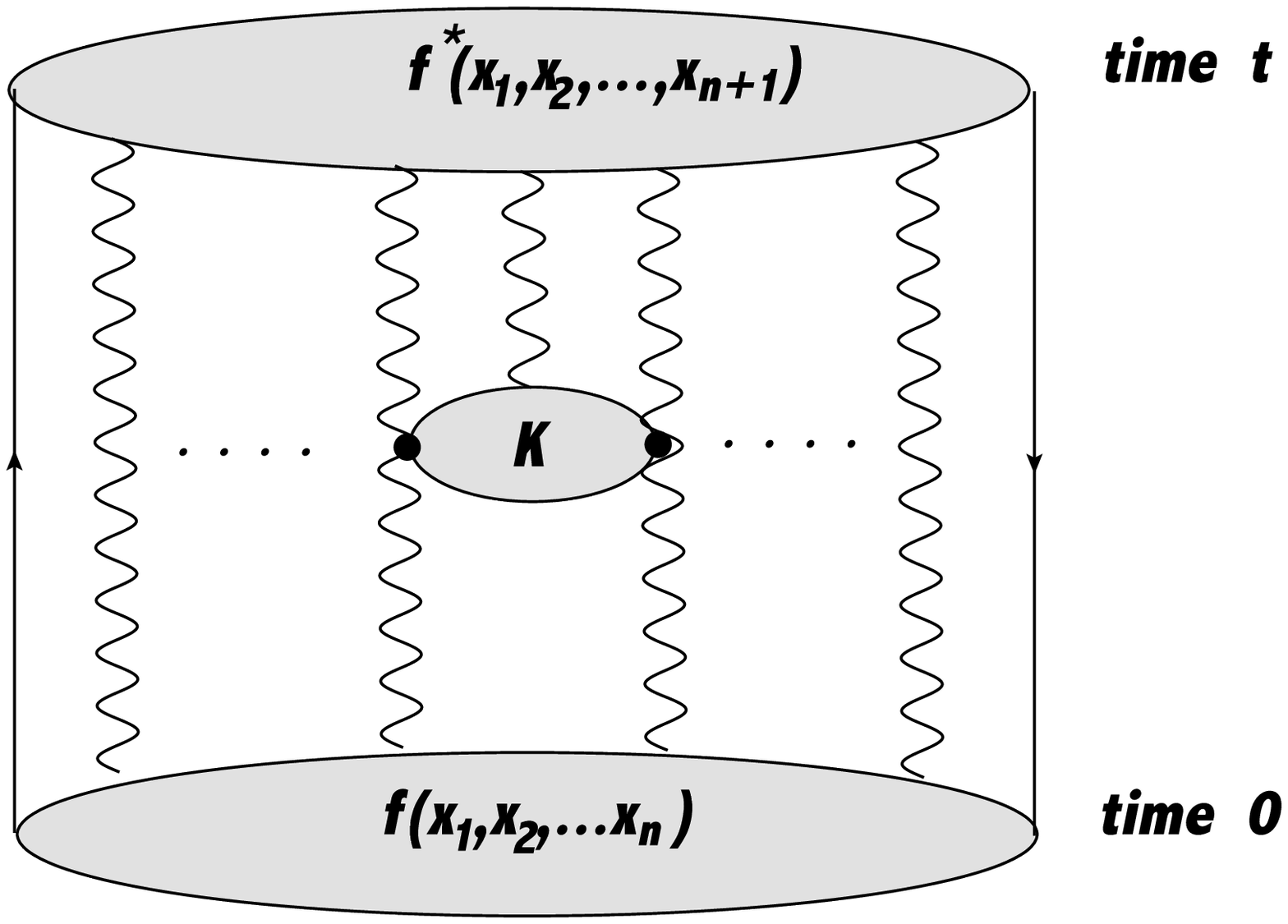}
}
\caption{Diagrams which, after a time derivative, contribute to Hamiltonian matrix elements.  (a) The graph which determines the kinetic energy of constituent gluons, contributing to $H_{nn}$. (b) One of the graphs determining the Coulomb energy of the 
$n$-gluon state, also contributing to $H_{nn}$.  The blob labeled ``K'' is the instantaneous gluon propagator $\langle K \rangle$. (c) Schematic of a graph contributing to an off-diagonal element $H_{n,n+1}$.  Here one of the $A$-field operators in the Coulomb operator 
$K(A)$ contracts with a gluon in the final state.}
\label{graphs}
\end{figure*}

   For the sake of simplicity, let us imagine that the $n$-gluon constituent states are already a set of
orthonormal states, i.e.\ $|\tn \rangle = |n\rangle_{\q q}$.  The diagrams contributing to $T_{nn}(t)$ which are responsible for the kinetic and Coulombic contributions to $H_{nn}$ are sketched in Figs.\ \ref{kinetic} and \ref{coulomb}.  Here the wavy lines are transverse gluon propagators.  The blob is the Coulomb propagator, i.e.\ the VEV of the $K$ operator.  At each end of this propagator one can attach either the fermion charge operator   $\rho^a_{q/\bar q}(\vx)$  (which in turn attaches to an external heavy quark or Wilson line), or the gluonic charge operator  $\rho^a_g(\vx)$  whose field operators connect to constituent gluons in the initial and final states, as indicated in the figure.

    The kinetic energy of the $n$-gluon state derives from the time derivative of the diagram in Fig.\ \ref{kinetic}.  A rough estimate
of this energy, for a wavefunction of the type shown in \rf{f}, goes as follows:  The uncertainty in position of the gluon along the
$z$-axis is approximately $R/n$, while the uncertainty in the transverse directions is $\sqrt{2}/a$.  For a massless gluon, ignoring modifications that might arise from the Gribov form of the propagator, the total kinetic energy is
\beq
           E_{kin}  =  n \sqrt{ {n^2 \over R^2} + a^2} \ .
\label{Ekin}
\eeq
In the Appendix we will explain the relationship between this estimate and $n$ particle state defined in (\ref{state1}-\ref{f}).

   The Coulomb energy due to interactions between $n$ nearest-neighbor gluons is proportional to their average separation, and originates from diagrams of the type shown in Fig.\ \ref{coulomb}.  For $n$ constituent gluons the average separation (ignoring transverse fluctuations) is roughly $R/(n+1)$, so the  Coulomb energy for a nearest-neighbor pair is about $\s_c R/(n+1)$. There are $n+1$ diagrams of the type in Fig.\ \ref{coulomb} (counting interactions with the external lines), so summing all inter-gluon separations we have $E_{Coul} \approx \s_c R$. Then we can estimate the diagonal matrix element as
\beq
           H_{nn} = n \sqrt{ {n^2 \over R^2} + a^2} + \s_c R \ .
\label{diagonal}
\eeq
We do not know much about the off-diagonal elements, except that, counting interactions with the external sources, there are 
$n+1$ diagrams of the form shown in Fig.\ \ref{offdiagonal}
which contribute to $H_{n,n+1}$.  However, each these diagrams is a function of the the average
gluon separation, and assuming some simple power dependence on average separation we would have, adding up all $n+1$
diagrams,
\beq
           H_{n,n+1} = H_{n+1,n} = (n+1) \a \left({R \over n+1}\right)^p \ ,
\label{offdiag}
\eeq 
where $\a$ is some dimensionful constant, and $p$ is an unknown (positive or negative) power.  We will neglect for now all other off-diagonal elements.  It will be convenient, for display purposes, to adopt units such that $\s_c=1$ in addition to the usual choice of $\hbar=c=1$.  We can now truncate
the basis and, for some choice of $\a,a,p$, extract the lowest eigenvalue of $H_{mn}$.  This is the potential $V(R)$ of the
lowest energy state available in the truncated basis.

\begin{figure}[t!]
\centerline{\scalebox{0.6}{\includegraphics{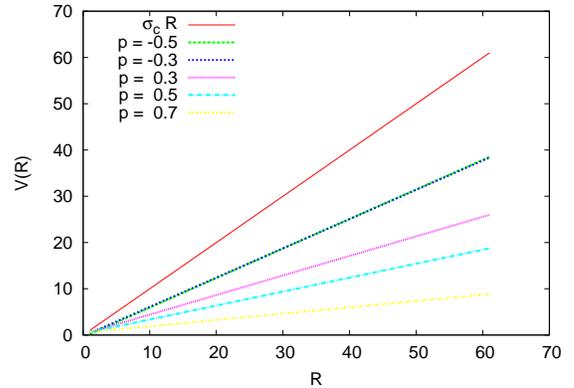}}}
\caption{The potential between static sources in the simple model outlined in the text, for fixed parameters $\a=0.7,~a=0.3$ and various powers $p$ in the off-diagonal Hamiltonian matrix elements.  Units are $\s_c=1$, and the upper line is the instantaneous Coulomb potential, which has slope=1 in these units.}
\label{model}
\end{figure}

\begin{figure}[t!]
\centerline{\scalebox{0.6}{\includegraphics{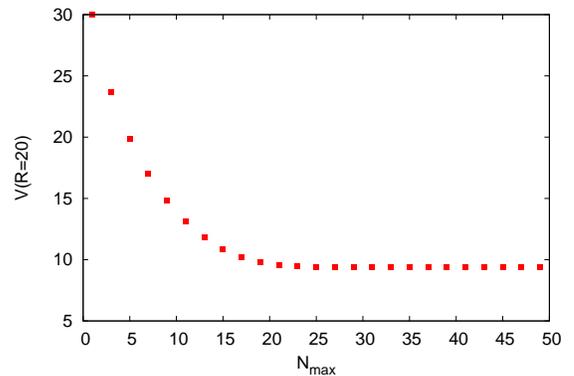}}}
\caption{Dependence of the static potential on number of $n$-gluon basis states in the model calculation.  Here $N_{cut}$ is the
maximum number of constituent gluons, $R=30, ~p=0.5$ and parameters $\a,a$ as before.}
\label{ncut}
\end{figure}  

    In Fig.\ \ref{model} we show the potential $V(R)$ which is obtained for a choice of parameters $a=0.3, \a=0.7$, and
a variety of powers $-0.5\le p \le 0.7$.   In this figure the top line, which has a slope = 1, is the non-perturbative Coulomb
potential $V(R)=\s_c R$ in units $\s_c=1$.   What is striking in this plot is that the potential for the lowest energy state
is linear in $R$, regardless of the power $p$, and even regardless of the overall sign of the off-diagonal element.  In addition, the string tension in each case is lower than $\s_c$ by factors ranging from 1.6 to 6.7.  The string tension for any $p$ can be adjusted by adjusting the parameters, but the point here is that the gluon chain result is robust: the linearity of the static quark potential, and the fact that the string tension can be much lower than the Coulomb string tension, seems to be generic in this setup, for a large range of power behavior in the off-diagonal term $H_{n,n+1}$.   For the data in the figure we have chosen to cut off the basis at $n=N_{cut}=100$, but in fact this is not necessary.  At the smaller values of $R$ a cutoff at much 
smaller $n$ will not change the results, and in general one can choose a cutoff which grows linearly with $R$.  This is illustrated
in Fig.\ \ref{ncut}, where we plot, at $R=30$ and $p=0.5$, the lowest energy eigenvalue of the truncated Hamiltonian matrix, 
$V(R)$, as a function of the truncation at $n=N_{cut}$ in the number of basis vectors $|n\rangle_{\q q}$

    Of course there is no guarantee that further off-diagonal terms, i.e..\ $H_{n,n+m}$ for $m\ge 2$, are negligible, and
it is not clear how such terms would change the picture.  Ultimately it will be necessary to estimate such terms, and we hope to return to this issue in a later publication.  But, at a minimum, we have seen that it is not difficult to understand how the asymptotic string tension could be several times smaller than the Coulomb string tension.  In fact, in the context
of the simple model presented above, this effect seems to be natural.

\section{Conclusions}

   We have shown that for SU(3) lattice pure-gauge theory the Coulomb potential in Coulomb gauge has the long-distance
behavior ($R$ in fm)
\beq 
          V_c(R) = \s_c R - (0.197 ~ \text{GeV-fm}){\g \over R} \ ,
\eeq
where 
\beq
          \s_c = 4.03(8) ~ \text{Gev/fm} = (891(9) ~\text{MeV})^2 \ ,
\eeq
which is a little more than four times the accepted value of (440 MeV)${}^2$ for the asymptotic string tension.  Our value
for $\s_c$ is taken from the data point at $\b=6.4, a=0.051$ fm, where the data appears, from Fig.\ \ref{tension}, to have converged to the $a \ra 0$ limit.  The dimensionless constant $\g$ seems consistent, for unknown reasons, with
the coefficient ${\pi \over 12}$ of the L\"uscher term.  We have also shown, in the context of a very simple
model based on the gluon chain picture, how the string tension of the lowest energy state with static $\q q$ sources, which we take to be the asymptotic string tension, can be so much smaller than the Coulomb string tension.  

    It would be interesting to attempt a more quantitative treatment of flux tube formation and, perhaps, heavy meson physics, using Coulomb, ghost, and gluon propagators taken from Monte Carlo simulations to estimate Hamiltonian matrix elements in a finite basis.  We hope to report on work along these lines at a later time.

\acknowledgments{This work was supported in part by the U.S. Department of Energy under Grants No. 
DE-FG03-92ER40711 (JG) and  DE-FG0287ER40365 (APS). The work was authored in part by Jefferson Science 
Associates, LLC under U.S. DOE Contract No. DE-AC05- 06OR23177.}

\vspace{20pt}

\appendix*

\section{}

   In this appendix we will explain the relationship between a gluon chain state (\ref{state1}-\ref{f}) and the rough estimate for
gluon kinetic energy \rf{Ekin}.  It will simplify the discussion to ignore vector and color indices as well as interactions, so let us consider a free massive scalar field with many flavors, with flavor 1 associated with position $x_1$,
flavor 2 associated with position $x_2$, and so on, and define the $N$-particle state
\beq
       |N\rangle =   \psi_N[\phi;f] |0\rangle_{vac} \ ,
\eeq
where       
\bea
 \psi_N[\phi;f] &=&        
       \int d^3x_1 d^3x_2 ... d^3x_n ~ f(\vx_1,\vx_2,...,\vx_N)
\non \\
  & &      \times \tp_1(\vx_1)
       \tp_2(\vx_2)... \tp_N(\vx_n)  \ .
\eea
 and where the subscripts denote flavors.  We assume the function $f$ is normalized, i.e.
\beq
            \int \left(\prod_{n=1}^N d^3x_n\right) f^*(x_1,x_2,...,x_N) f(x_1,x_2,...,x_N) = 1
\eeq
The scalar field operators are $\phi_n(x)$, or $\phi_n(k)$ in momentum space, and
\beq
\tp_n(x) \equiv \int {d^3k \over (2\pi)^3} ~ \sqrt{2 \o_k} \phi_n(k)
\eeq
with $\o_k=\sqrt{k^2+m^2}$.  Propagators are
\bea
          D_{ij}(\vx-\vy,t) &=& \langle \phi_i(x,t) \phi_j(y,0) \rangle = \d_{ij} D(\vx-\vy)
\non \\
&=& \d_{ij} \int {d^3k \over (2\pi)^3} ~ {e^{i\vk \cdot (\vx-\vy)} e^{-\o_k t} \over 2\o_k}
\eea
and
\bea
\tD_{ij}(\vx-\vy,t) &=& \langle \tp_i(x,t) \tp_j(y,0) \rangle
\non \\
 &=& \d_{ij} \int {d^3k \over (2\pi)^3} ~ e^{i\vk \cdot (\vx-\vy)} e^{-\o_k t}
\eea
If $f^{(N)}$ is normalized, then so is $|N\rangle$:
\begin{widetext}
\bea
\langle N | N \rangle &=& \int \left\{\prod_{n=1}^N d^3x'_n d^3x_n\right\}   f^*(x'_1,...,x'_N) f(x_1,...,x_N)
\non \\
& & \qquad \times  \langle \tp_1(\vx'_1,0) \tp_2(\vx'_2,0)...\tp_N(\vx'_N,0) \tp_1(\vx_1,0)
 \tp_2(\vx_2,0)...\tp_N(\vx_N,0) \rangle
\non \\
&=& \int \left\{\prod_{n=1}^N d^3x'_n d^3x_n\right\}   f^*(x'_1,...,x'_N) f(x_1,...,x_N) \tD(\vx'_1-\vx_1,0) 
\tD(\vx'_2-\vx_2,0)...\tD(\vx'_N-\vx_N,0)
\non \\
&=&  \int \left\{\prod_{n=1}^N d^3x_n\right\}   f^*(x_1,...,x_N) f(x_1,...,x_N)
\non \\
&=& 1
\eea
Then, to compute the energy expectation value 
\bea
\langle N|H|N \rangle &=&  -\lim_{t\ra 0} {d \over dt} \langle N|e^{-H t}|N \rangle
\non \\
&=&  -\lim_{t\ra 0} {d \over dt} \int \left\{\prod_{n=1}^N d^3x'_n d^3x_n\right\}   f^*(x'_1,...,x'_N) f(x_1,...,x_N)
 \langle \tp_1(x'_1,t) \tp_2(\vx'_2,t)...\tp_N(\vx'_N,t) \tp_1(\vx_1,0) \tp_2(\vx_2,0)...\tp_N(\vx_N,0) \rangle
\non \\
&=&  -\lim_{t\ra 0} {d \over dt} \int \left\{\prod_{n=1}^N d^3x'_n d^3x_n\right\}   f^*(x'_1,...,x'_N) f(x_1,...,x_N) 
  \tD(\vx'_1-\vx_1,t) \tD(\vx'_2-\vx_2,t)...\tD(\vx'_N-\vx_N,t)
\non \\
&=&    \int \left(\prod_{i=1}^N {d^3k_l \over (2\pi)^3} \right) \left(\sum_n \o_{k_n}\right) F^*(k_1,...,k_N) F(k_1,...,k_N)
\non \\
&=& \left\langle \! \! \! \left\langle \sum_{n=1}^N \sqrt{k_{nx}^2 + k_{ny}^2 + k_{nz}^2 + m^2} \right\rangle \! \! \! \right\rangle
\eea
\end{widetext}
where $F(k_1...k_N)$ is the Fourier transform of $f(x_1...x_N)$, and the $\langle \! \langle ... \rangle \! \rangle$ symbol indicates
an ordinary quantum mechanics expectation value in the $N$-particle wavefunction specified by $f$.  

   A first approximation is to take the expectation values inside the square root
\bea
\langle N|H|N \rangle &\approx& \sum_{n=1}^N \sqrt{\langle \! \langle k_{nx}^2 \rangle \! \rangle + \langle \! \langle k_{ny}^2 \rangle \! \rangle + \langle \! \langle k_{nz}^2\rangle \! \rangle + m^2 }
\non \\
&=& \sum_{n=1}^N \sqrt{\D k_{nx}^2 + \D k_{ny}^2 + \D k_{nz}^2 + m^2} 
\eea
Applying the approximate relation $\D k_{nx} \approx 1/\D x_n$, where $\D x_n$ is the positional uncertainty of
the $n$-th particle in wavefunction $f$, we can estimate that
\beq
\langle N|H|N \rangle \approx  \sum_{n=1}^N \sqrt{ {1\over \D x_n^2} +{1\over \D y_n^2} +{1\over \D z_n^2} +m^2}
\eeq 
Finally, if $f$ represents a chain state analogous to \rf{f},  then transverse fluctuations $\D x_n = \D y_n = \r$ are approximately the same for each of the $N$ constituent particles, and $\D z_n \approx R/N$.   Then we have
\beq
\langle N |H| N \rangle \approx N \sqrt{ {N^2 \over R^2} + {2\over \r^2} + m^2}
\eeq
For a massless particle, this is the kinetic energy estimate given in \rf{Ekin}.

\bibliography{chain}

\end{document}